\documentclass[useAMS,usenatbib]{mn2e}
\usepackage{graphicx}    
\DeclareGraphicsExtensions{.ps,.pdf,.png}
\usepackage{natbib}
\makeatletter
\def\fps@figure{htbp}
\makeatother

\begin{document}
\title[Massive Galaxy Morphology]{The Redshift and Mass Dependence on the Formation of The Hubble Sequence at $z>1$ from CANDELS/UDS}
\author[Mortlock et al.]{Alice~Mortlock$^{1}$, Christopher~J.~Conselice$^1$, William G. Hartley$^1$, Jamie R. Ownsworth$^{1}$, 
\newauthor Caterina Lani$^{1}$, Asa F. L. Bluck$^{2}$, Omar Almaini$^{1}$, Kenneth Duncan$^{1}$,  Arjen van der Wel$^{3}$,
\newauthor Anton M. Koekemoer$^{4}$, Avishai Dekel$^{5}$, Romeel Dav\'{e}$^{6}$, Harry C. Ferguson$^{4}$,
\newauthor Duilia F. de Mello$^{7}$, Jeffrey A. Newman$^{8}$, Sandra M. Faber$^{9}$, Norman A. Grogin$^{4}$,
\newauthor Dale D. Kocevski$^{10}$, Kamson Lai$^{9}$
\footnotemark[0]\\ $^{1}$University of Nottingham, School of Physics and Astronomy, Nottingham, NG7 2RD UK \\ $^{2}$University of Victoria,
3800 Finnerty Road, Victoria BC, V8P 5C2, Canada \\ $^{3}$Max-Planck Institut f\"ur Astronomie, K\"onigstuhl 17, D-69117, Heidelberg, Germany \\ $^{4}$Space Telescope Science Institute, 3700 San Martin Drive, Baltimore, MD 21218, USA \\ $^{5}$Racah Institute of Physics, The Hebrew University, Jerusalem 91904 Israel \\ $^{6}$University of Arizona, 933 N. Cherry Ave, Tucson, AZ 85721, USA \\ $^{7}$Physics Department, Catholic University of America, 200 Hannan Hall,Washington, DC 20064, USA \\ $^{8}$Department of Physics and Astronomy, University of Pittsburgh, Pittsburgh, PA 15260, USA \\ $^{9}$UCO/Lick Observatory, Department of Astronomy and Astrophysics, University of California, Santa Cruz, CA 95064, USA \\ $^{10}$Department of Physics and Astronomy, University of Kentucky, Lexington, KY 40506, USA}

\date{Accepted ??. Received ??; in original form ??}
\pagerange{\pageref{firstpage}--\pageref{lastpage}} \pubyear{2002}
\maketitle

\label{firstpage}

\begin{abstract}
In this paper we present a detailed study of the structures and morphologies of a sample of 1188 massive galaxies with $M_{*}\ge10^{10}M_{\odot}$ between redshifts $z=1-3$ within the Ultra Deep Survey (UDS) region of the Cosmic Assembly Near-infrared Deep Extragalactic Legacy Survey (CANDELS) field. Using this sample we determine how galaxy structure and morphology evolve with time, and investigate the nature of galaxy structure at high redshift. We visually classify our sample into disks, ellipticals and peculiar systems and correct for redshift effects on these classifications through simulations. We find significant evolution in the fractions of galaxies at a given visual classification as a function of redshift. The peculiar population is dominant at $z>2$ with a substantial spheroid population, and a negligible disk population. We compute the transition redshift, $z_{trans}$, where the combined fraction of spheroidal and disk galaxies is equal to that of the peculiar population, as $z_{trans}=1.86\pm 0.62$ for galaxies in our stellar mass range.  We find that this transition changes as a function of stellar mass, with Hubble-type galaxies becoming dominant at higher redshifts for higher mass galaxies ($z_{trans}=2.22\pm0.82$), than for the lower mass galaxies ($z_{trans}=1.73\pm0.57$). Higher mass galaxies become morphologically settled before their lower mass counterparts, a form of morphological downsizing. We furthermore compare our visual classifications with S\'{e}rsic index, the concentration, asymmetry and clumpiness (CAS) parameters, star formation rate and rest frame $U-B$ colour. We find links between the colour of a galaxy, its star formation rate and how extended or peculiar it appears. Finally, we discuss the negligible $z>2$ disk fraction based on visual morphologies and speculate that this is an effect of forming disks appearing peculiar through processes such as violent disk instabilities or mergers. We conclude that to properly define and measure high redshift morphology and structure a new and more exact classification scheme is needed.

\end{abstract}

\begin{keywords}
galaxies: evolution--galaxies: formation--galaxies: structure--galaxies: general
\end{keywords}

\newpage

\section{Introduction}
\label{sec:int}
In the local Universe the visual morphologies of galaxies are well described by the Hubble tuning fork diagram (\citealt{Hubb26}). Galaxies tend to be either smooth and elliptical or disk-like, sometimes with features such as spiral arms and bars. These two classes of galaxies are well studied, and there are well defined trends associated with them. Elliptical galaxies are often red in colour and are non-starforming systems, while galaxies with disk morphologies are bluer with higher star formation rates (e.g. \citealt{Sand86}, \citealt{Stra01}, \citealt{Kauf03}, \citealt{Cons06}, \citealt{Nair10}). Structurally, elliptical galaxies have higher concentrations, lower asymmetries and higher S\'ersic indices than the low redshift disk population (e.g \citealt{Cons03}, \citealt{Cons05}, \citealt{Scar07}, \citealt{Nair10}).

Studies of galaxies around redshift $z=1$ find galaxies with similar morphologies to those at low redshift (e.g \citealt{Brin98}, \citealt{Abra96}, \citealt{Vand00}, \citealt{Cons05}, \citealt{Papo05}, \citealt{Ilbe06}, \citealt{Oesc10} and \citealt{Buit13}). These studies find that at $z<1$ the population of irregular galaxies is similar to the population found today and that the Hubble sequence galaxies dominate. Further to this, many of the local relationships between structural parameters and galaxy morphology are still present. For example, there are still links between CAS morphology and visual morphology at $z\sim1$ (\citealt{Cons05}, \citealt{Bluc12}) and there are relations between colour, star-formation rate and visual morphology at this redshift (e.g. \citealt{Bell04}). 

However, the picture of galaxy morphology at higher redshift is less clear. Some studies (e.g. \citealt{Dick00}, \citealt{Papo05}, \citealt{Came11}) find that there are almost no Hubble type galaxies present at $z>2$ and hence there must be large amounts of evolution occurring to transform the irregular galaxies seen in the high redshift Universe into their more settled counterparts that we see today. Conversely other studies (e.g. \citealt{Driv98}, \citealt{Cons05} \citealt{Szom11}, \citealt{Cons11} and \citealt{Buit13}) find a dominant peculiar population, but also, that normal Hubble galaxies that we find in the local Universe do in fact exist at $z>2$.

Further to the different results regarding the presence of the Hubble sequence, there is disagreement amongst exactly which population dominates at high redshift. Some studies find that there is a large irregular population and that Hubble type galaxies emerges somewhere between $z=1-3$ (\citealt{Krie09} and \citealt{Szom11}), whereas others find evidence that, whilst there is a strong peculiar population, there is also a large visually disk-like or elliptical population at high redshift (\citealt{Cons11} and \citealt{Buit13}). There could be several reasons for these discrepancies, such as variations in completeness or sample selection or effects of image quality on classifications. It is clear that further work on resolving these differences is needed, and for that a large sample of galaxies with detailed high resolution rest-frame optical imaging is required. By exploring the morphologies of galaxies in the optical we are not biased towards very blue features, a problem which could be part of the explanation for the differences in studies so far. 

Overall, we know that a galaxy's structure is linked with the processes that occur in its lifetime. Therefore, if we are to fully understand galaxy evolution we need to understand galaxy morphology. The Cosmic Assembly Near-infrared Deep Extragalactic Legacy Survey (CANDELS) (\citealt{Grog11} and \citealt{Koek11}) provides very deep, high resolution, near infrared imaging of some of the most well covered areas of the sky. The resolution of the WFC3 camera and the depth of CANDELS is vital for obtaining reliable visual morphologies. Furthermore, the $H_{160}$ imaging probes optical light in the redshift range $z=1-3$, making this data ideal for studying the visual morphologies of high redshift galaxies, unlike several past surveys which have imaged galaxies in the rest-frame UV. Rest-frame UV light is dominated by star formation features which may not represent well the underlying light distributions.

There are already several studies using the CANDELS data investigating the morphologies of different samples of galaxies (e.g. \citealt{Bruc12} \citealt{Kart12}, \citealt{Koce12}, \citealt{Tarr13}). \citet{Kart12}, \citet{Koce12}, \citet{Tarr13} investigate the visual morphologies of Ultraluminous Infrared Galaxies (ULIRGS), AGN and sub-mm galaxies respectively. They find that AGN and sub-mm galaxies are generally relaxed/normal systems, whereas ULIRGS often show signs of mergers/peculiarities. \citet{Bruc12} look at the bulge to disk decomposition of massive galaxies ($M_{*}\ge10^{11}M_{\odot}$) and show that their sample is dominated by galaxies with disk morphologies in the redshift range $1<z<3$. 

Further CANDELS studies have explored the structure and physical properties of galaxies (e.g. \citealt{Bell12}, \citealt{Wang12}, \citealt{Barr13}). These works have found links between S\'ersic index, stellar mass, size, structure and star formation history at redshifts of $z>1$ for different galaxy samples. This paper adds to these studies by looking at the visual and structural morphologies of a $M_{*}\ge10^{10}M_{\odot}$ and $z=1-3$ sample of galaxies in the UDS part of CANDELS, as a function of redshift and mass, and how these morphologies compare to various physical parameters. We take advantage of the full, multi-wavelength, UDS data set to obtain redshifts, stellar masses and other important galaxy properties. To visually classify these galaxies we use the $H_{160}-$band CANDELS UDS image, which provides us with deep (5$\sigma$ point source depth of H=27.0 mag), rest-frame optical imaging over 0.06 arcminutes$^2$ of the full UDS field.

The paper is set out as follows. Section \ref{sec:data} describes how we calculate redshifts, stellar masses, rest frame colours, CAS parameters and star formation rate. Section \ref{sec:visclass} describes our sample selection, visual classification system, and caveats in the system. Section \ref{sec:results} describes our results and Section \ref{sec:diss} provides a discussion of these results. Throughout this paper we assume $\Omega_{M}=0.3$, $\Omega_{\Lambda}=0.7$ and $H_{0}=70$ km s$^{-1}$ Mpc$^{-1}$. AB magnitudes and a Chabrier initial mass function (IMF) are used throughout.

\section{Data and Sample}
\label{sec:data}
In this work we choose a sample of $z=1-3$ galaxies with $M_{*}\ge10^{10}M_{\odot}$. The redshifts and stellar masses are computed using ground based UDS data (Almaini et al., in preparation) as described in Section \ref{sec:zandm}. From this we obtain a sample of 1213 massive, high redshift galaxies which we visually classify. To perform our visual classifications we take advantage of the space based CANDELS (PIs Faber/Ferguson) $H_{160}-$band data.

The ground based data is from the UKIRT Infrared Deep Sky Survey (UKIDSS, \citealt{Lawr07}) Ultra Deep Survey (UDS) DR8 data release, and reaches 5$\sigma$, 2$^{"}$-aperture depths of J=24.9, H=24.2 and K=24.6. The UDS covers a total of 0.88 deg$^2$ and has additional wavelength coverage from various other surveys: (Optical data from the Subaru-XMM Deep Survey (\citealt{Furu08}), Infrared data from the Spitzer Legacy Program (PI:Dunlop) and $U-$band data from the Canada France Hawaii Telescope (CFHT; Foucaud et al. in preparation). For further information on the UDS see Almaini et al. (in preparation).

CANDELS is an on going Multi Cycle Treasury Programme consisting of 902 orbits of the HST. It utilises two HST cameras, the Advanced Camera for Surveys (ACS) and the Wide Field Camera 3 (WFC3), and when complete it will have imaged over 250,000 galaxies at $z<10$. CANDELS covers roughly 800 arcminutes$^2$ in total, comprising of five different fields: GOODS-N, GOODS-S, EGS, COSMOS and UDS. This is further split into two parts, CANDELS/Deep which images GOODS-N and GOODS-S to a 5$\sigma$ point source depth of H=27.7 mag, and CANDELS/Wide which images all fields to a 5$\sigma$ point source depth of H=27.0 mag. In this study, we perform our visual classification on the CANDELS data that covers a part of the UDS field specifically. The area covered by this part of the survey is $9.4^{"} \times 22.0^{"}$, is at two orbits depth in the $H$ and $J-$bands and has a pixel scale of $0.06$ arcseconds/pixel. For further details on the CANDELS data see \citet{Grog11} and \citet{Koek11}. For a detailed discussion of the CANDELS UDS field see \citet{Gala13}.

\subsection{Redshifts, Stellar Masses and Rest Frame Magnitudes}
\label{sec:zandm}
\begin{figure}
\centering
\includegraphics[trim = 4mm 0mm 0mm 6mm, clip,scale=0.6]{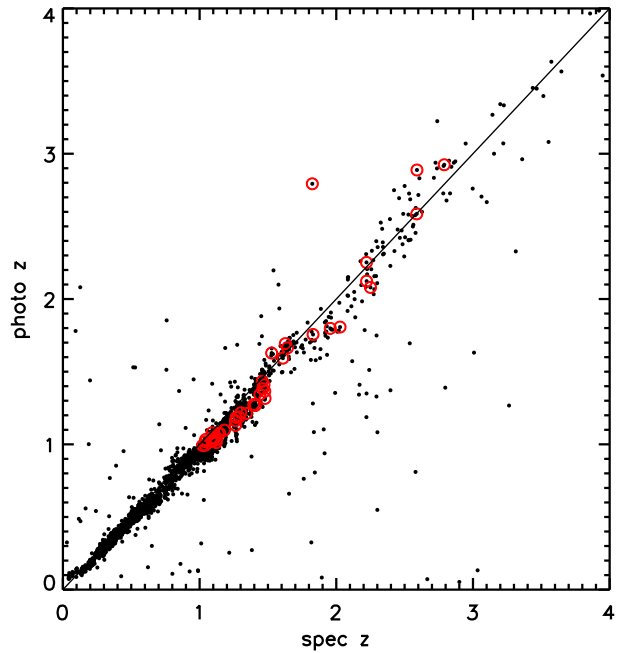}
\caption{Spectroscopic redshifts versus photometric redshifts for 2146 galaxies in the UDS. The red circles highlight the 44 massive galaxies with spectroscopic redshifts used in this work.}
\label{speczphotoz}
\end{figure}
Photometric redshifts are determined via fitting template spectral energy distributions (SEDs) to the $UBVRizJHK$ and Infrared Array Camera (IRAC) Channel 1 and 2 photometric data points, and are computed using the \textsc{EAZY} code (\citealt{Bram08}) and include an apparent $K-$band magnitude prior. The photometry is fit to the six default \textsc{EAZY} templates, and an additional template which is the bluest \textsc{EAZY} template with a small amount of SMC-like extinction added. The redshifts are computed using a maximum likelihood analysis. For full details of the fitting procedure and resulting photometric redshifts see \citet{Hart13}.

We compare the photometric redshifts used in this work to spectroscopic redshifts that are available in the UDS. Of the available spectroscopic redshifts, 1500 are from the UDSz, an ESO large spectroscopic survey (ID:180.A-0776) within the UDS. The UDSz imaged a large sample of galaxies at $z>1$ with $K<23.0$ (for this study the photometric redshifts are measured to a depth of $K\sim24.5$). A further 4000 archival redshifts were taken from the literature, see \citet{Simp12} and references therein for a detailed description of these spectroscopic redshifts. We therefore have a total of $\sim$5500 spectroscopic redshift, however, this number is reduced to 2146 after the removal of AGN. This was done by removing both X-ray and radio sources, and also, by removing objects which had AGN signatures in their spectra. Excluding catastrophic outliers ($\Delta z/(1+z)>0.15$), we find the dispersion of $z_{photo} - z_{spec}$ is $\Delta z/(1+z)=0.031$. Figure \ref{speczphotoz} shows the spectroscopic redshifts versus the photometric redshifts for the 2146 galaxies used to calculate the dispersion. Circled in red are the massive galaxies used in this study. We find that $\Delta z/(1+z)=0.031$ for the galaxies circled in red. For the 44 massive galaxies used in this study, 68\% of the spectroscopic redshifts are below a redshift of $z=1.5$ (23\% are around a redshift of $z\sim1$). Also, we note that the average of the 44 spectroscopic redshifts is slightly higher than that of their corresponding photometric redshifts ($\langle z \rangle=1.48$ for the spectroscopic redshifts compared to $\langle z \rangle=1.42$ for the photometric redshifts). However, when we take into account the photometric errors these two are comparable.

The stellar masses and rest frame magnitudes used in this work are measured using a multicolour stellar population fitting technique where we fit to the $UBVRizJHK$ bands and IRAC Channel 1 and 2 bands. A large grid of synthetic SEDs are constructed from the stellar population models of \citeauthor{Bruz03} (2003, hereafter BC03), assuming a Chabrier IMF (\citealt{Chab03}). The star formation history is characterised by an exponentially declining model with various ages, metalicities and dust extinctions.
These models are parametrised by an age of the onset of star formation, and by an e-folding time such that
\begin{equation}
SFR(t) \sim SFR_{0}\times \rm e ^{-\frac{t}{\tau}}.
\label{eq:SFRexp}
\end{equation}
\noindent where the values of $\tau$ ranges between 0.01 and 10.0 Gyr, while the age of the onset of star formation ranges from 0.001 to 13.7 Gyr. The metalicity ranges from 0.0001 to 0.05 (BC03), and the dust content is parametrised by $\tau_{v}$, the effective $V-$band optical depth for which we use values $\tau_{v}$ = 0, 0.4, 0.8, 1.0, 1.33, 1.66, 2, 2.5, 5.0. We do not investigate other star formation histories in this work, however, studies have shown that stellar mass calculations are generally robust to changes in star formation history within our redshift range and for the stellar masses we probe (e.g. \citealt{Ilbe10}, \citealt{Owns12}, \citealt{Pfor12}, \citealt{Ilbe13}).

To fit the SEDs we first scale them to the apparent $K-$band magnitude of the galaxy we are fitting. We then fit each scaled model template in the grid of SEDs to the measured photometry of the galaxy. We compute the $\chi^{2}$ values for each template and select the best fitting one. From this we obtain a best fit stellar mass and best fit rest frame magnitudes. We also calculate a modal mass value by binning the stellar masses of the ten percent of templates with the lowest $\chi^{2}$ in bins of 0.05 dex. The mode stellar mass corresponds to the stellar mass bin with the largest number of templates. In this analysis we use the mode stellar masses and the best fit rest frame magnitudes. We note that we do not take into account the errors on the photometric redshifts when calculating the stellar masses, however, these errors are included in all further error analysis in this work (see Section \ref{sec:errors}).

\begin{figure}
\centering
\includegraphics[trim = 4mm 0mm 0mm 6mm, clip,scale=0.6]{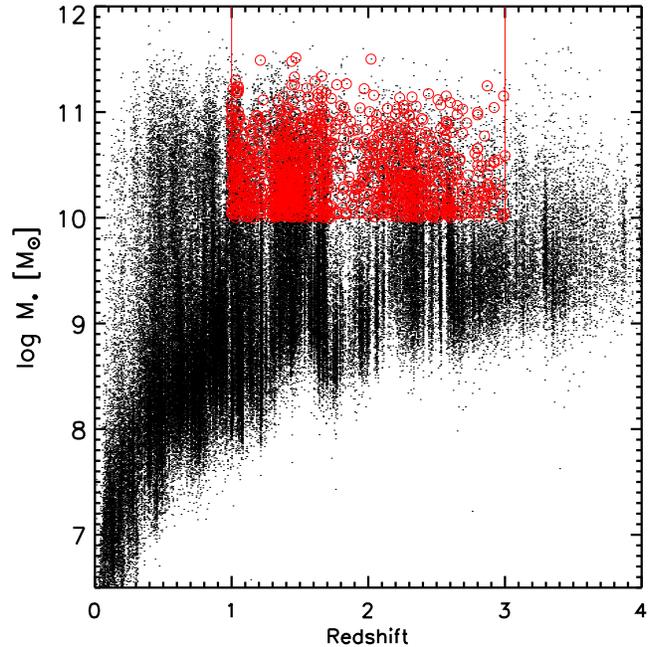}
\caption{The stellar masses as a function of redshift for all galaxies in the UDS (black dots). The red circles indicate the 1188 $M_{*}\ge10^{10}M_{\odot}$ and $z=1-3$ galaxies used in this work taken from the CANDELS sub-region in the UDS.}
\label{redshiftmass}
\end{figure}

We choose mode stellar masses as these are less likely to be affected by templates which are formally the best-fit but may lead to erroneous stellar masses. For example, a template may be an extremely good fit to some of the photometry, resulting in a low $\chi^{2}$, but not accurately represent the overall galaxy photometry. We cannot use a modal rest-frame magnitude, however, as the greater number of templates with blue colours and their limited dynamic range in colour almost always results in a blue modal rest-frame colour. Therefore in this work we use the best fit rest frame magnitudes. The stellar masses as a function of redshift are shown in Figure \ref{redshiftmass} for all galaxies in the UDS (black dots) and the sample used in this work taken from the CANDELS part of the UDS (red circles).

The stellar mass completeness of the UDS is discussed in detail in \citet{Hart13}. This work shows that in the redshift range $z=2.5-3$ the UDS is complete down to a stellar mass of $M_{*}\sim10^{10}M_{\odot}$, even for red galaxies. Furthermore, in Mortlock et al. (in preparation) we show from simulations that we are $\sim100\%$ complete down to the $K-$band magnitudes used in this work. Therefore we are confident our stellar mass cut of  $M_{*}\sim10^{10}M_{\odot}$ gives us a sample which is complete in stellar mass for the full galaxy population.

\subsection{Star Formation Rates}
The $2800$\AA, UV star formation rates ($SFR_{2800}$) used in this paper are measured from rest-frame near UV luminosities. We determine then $SFR_{2800,uncorrected}$ for $z = 1.5-3$ galaxies from the observed optical Subaru $z-$band flux density. We determine the $SFR_{2800,uncorrected}$ for $z = 1-1.5$ galaxies from the optical Subaru $i-$band. After we apply an SED based k\--correction using the IDL \textsc{kcorrect} package (\citealt{Blan07}) fluxes correspond to a rest frame wavelength of $\sim2800$\AA.

We convert the UV luminosities to star formation rates using the \cite{kenn98} conversion from $2800$\AA$~$assuming a Chabrier IMF: 
\begin{equation}
SFR_{UV} (M_{\odot}\mathrm{yr^{-1}}) = 8.24 \times 10^{-29} L_{2800}
(\mathrm{ergs\, s^{-1}\, Hz^{-1})}
\end{equation}
 \noindent
The errors quoted here take into account photometric errors and the conversion from a luminosity. The error for individual star formation rates are around 30\%. This error is dominated by the dust correction discussed below.

UV light is very susceptible to dust extinction and a careful dust-correction has to be applied. To do this we compute the UV slope ($\beta$) from the best fit template SED. At the redshift ranges we probe we have many photometric data points in the UV and hence this part of the SED is well constrained.

We apply the method from \citet{Meur99} and the dust model from \citet{Fisc05} for determining a UV dust attenuation, $A_{2800}$, in terms of $\beta$. The equation used is:
\begin{equation}
A_{2800} = 1.67\beta+3.71
\end{equation}
\noindent where $A_{2800}$ is the amount of light lost due to dust in magnitudes. In recent studies (e.g. \citealt{Wije12}) it has been shown that this dust model is better suited to the general population of galaxies than the dust model of \citet{Calz00}. Although the dust model of \citet{Calz00} is often applied to galaxies with a large range in properties, the model itself is better suited to highly star-forming systems. We find that our dust corrections are in the range $A_{2800}=0-5$, and we add this too our $SFR_{2800,uncorrected}$ to obtain our final dust corrected star-formation rates.

We note that for passive galaxies the shape of the UV slope is not due to the presence of dust, but due to the presence of old stellar populations. Therefore, a dust correction calculated using the method described here will be incorrect, and hence result in an incorrect star-formation rate. To avoid this, we select a passive population of galaxies based on the $UVJ$ diagram as described in \citet{Hart13} and apply no dust correction to these galaxies. For a full description of the calculation of these star-formation rates see Ownsworth et al. (in preparation).

\subsection{CAS}
\label{sec:cas}
The CAS (concentration, asymmetry and clumpiness) parameters are a useful tool when investigating the morphologies of galaxies. For example, a highly asymmetric galaxy would be expected to have a peculiar visual morphology, and a galaxy with high concentration would be expected to have an early type morphology (\citealt{Cons03}). For this study we compute the CAS parameters using the CANDELS UDS $H_{160}-$band image. We do not use the clumpiness parameter as this is found to be the least robust at high redshift due to issues resolving small clumps in these systems with WFC3 (\citealt{Cons03}). Here we will include a brief discussion of how asymmetry and concentration are calculated. For an indepth discussion of how the CAS parameters are computed, including centering, measurement of radii and background subtraction see \citet{Cons00} and \citet{Cons03}.

The asymmetry parameter is found by subtracting a 180$^{\circ}$ rotated image of the galaxy from the original image. A background subtraction is included. The equation for this is as follows
\begin{equation}
A = min \left(\frac{\sum |I_{0}-I_{180}|}{\sum I_{0}}\right) - min \left(\frac{\sum |B_{0}-B_{180}|}{\sum I_{0}}\right)
\label{eq:asym}
\end{equation} 
where $I_{0}$ is the original image pixels and $I_{180}$ is the image after the 180$^{\circ}$ rotation. $B_{0}$ and $B_{180}$ are the values used for the background subtraction.

The concentration parameter is a measure of how concentrated the light is in a central region compared to a larger, less concentrated region. Mathematically we use
\begin{equation}
C = 5 \times  log \left(\frac{r_{80}}{r_{20}}\right)
\label{eq:conc}
\end{equation} 
where $r_{80}$ and $r_{20}$ are radii containing 80\% and 20\% of the galaxies total light respectively (\citealt{Bers00}).

Uncertainties on the asymmetry are the standard deviation of the background subtraction used in CAS along with photon counting errors from the galaxy itself. The uncertainties on concentration are propagated from the measurements of the radii. For a full discussion of how these errors are calculated see \citet{Cons00} and \citet{Cons03}.

\section{Visual Classification of the Sample}
\label{sec:visclass}

\subsection{The Classification System}
\label{sec:system}
\begin{figure*}
\centering
\includegraphics[trim = 20mm 99mm 0mm 20mm, clip,scale=1.15]{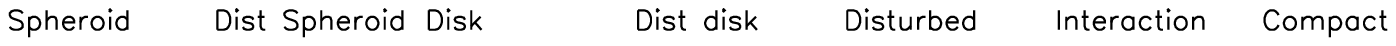}
\hrule
\includegraphics[trim = 20mm 10mm 0mm 27mm, clip,scale=1.15]{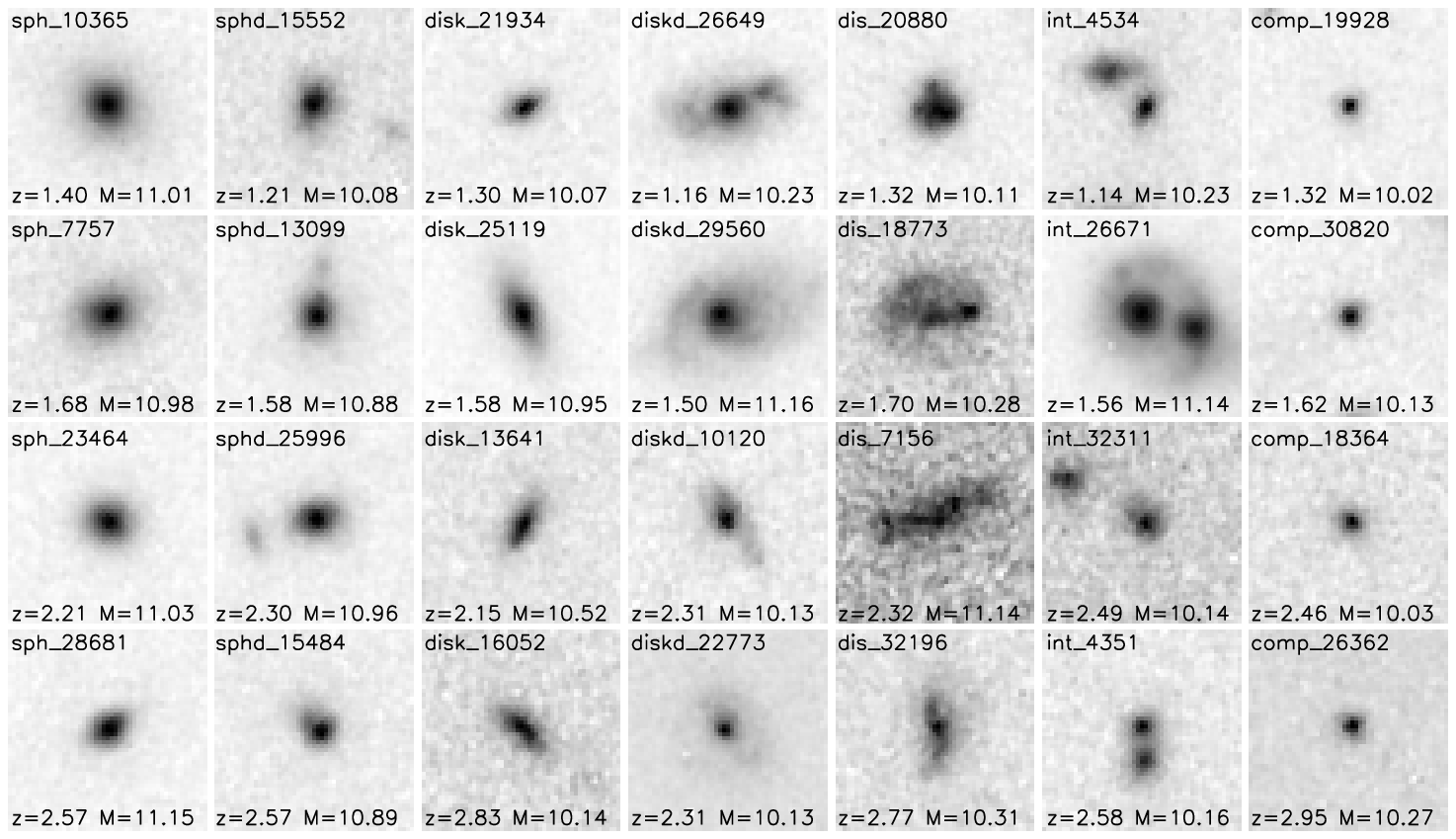}

\caption{Examples of the galaxies that fall into our various classification types. Each row is ordered by type. From left to right the order is spheroid (sph), disturbed spheroid (sphd), disk (disk), disturbed disk (diskd), disturbed (dis), interaction (int) and compact (comp). Each column is ordered by redshift bin. From top to bottom the order is $z=1-1.5$, $z=1.5-2$, $z=2-2.5$, $z=2.5-3$. The postage stamps are cut out from the CANDELS UDS $H_{160}-$band image and are approximately $3 \times 3$ arcseconds in size.}
\label{types}
\end{figure*}

Using the redshifts and stellar masses described in Section \ref{sec:data}, we define a sample of galaxies with $z=1-3$ and $M_{*}\ge10^{10}M_{\odot}$. We find 1213 galaxies that fall within these criteria. We reject 25 of these galaxies due to WFC3 image quality problems, thus we are left with a sample of 1188 galaxies which we visually classify using the $H_{160}-$band imaging. There are 9 categories that we use for our visual classifications, which are defined as follows.
\begin{itemize}

\item Type 0: Unclassifiable. Galaxies in this category are too small or too faint to classify.

\item Type 1: Spheroid. These galaxies are centrally concentrated, with a smooth profile and are roughly round/elliptical.

\item Type 2: Spheroid and disturbed. These galaxies are spheroidal, like the Type 1 galaxies, but also show some weak signs of peculiarity. However the dominant morphology of the galaxy is spheroid. 

\item Type 3: Disk. In this category galaxies show a disk in the form of an outer area of lower surface brightness than the central part of the galaxy. The disk part of the galaxy may or may not contain structure such as spiral arms.

\item Type 4: Disk and disturbed. These galaxies are the same as Type 3 galaxies but with some sort of disturbance, such as asymmetric spiral arms. However the disturbance is not large enough to destroy the overall disk morphology. 

\item Type 5: Disturbed. Any galaxy whose morphology is dominated by a disturbance or peculiarity and has no obvious disk or spheroid component.

\item Type 6: Interaction. In this category a galaxy must have a visually close companion that is approximately the same size as the galaxy being classified.

\item Type 7: Compact. Galaxies in this category appear to have small radii and spheroidal/smooth morphologies.

\item Type 8: Star or image problem. If the object is actually a star or there is some problem with the image (e.g. galaxy is close to the image edge) they are classed as Type 8 and discarded.

An example of some of the main types in our classification system can be seen in Figure \ref{types}.
\end{itemize}

When visually classifying galaxies there is a degree of subjectivity that affects the results. The method we use to try to limit the effects of this subjectivity is to have as many trained people classifying the galaxies as possible and then to use each individual opinion as a ``vote". In this work we have five classifiers, and we first look at galaxies where three or more classifiers agree. We find that there are 886 (75\%) galaxies which satisfy this criterion, and 302 (25\%) galaxies that are unclassified.

\begin{figure*}
\centering
\includegraphics[trim = 10mm 0mm 0mm 6mm, clip,scale=0.45]{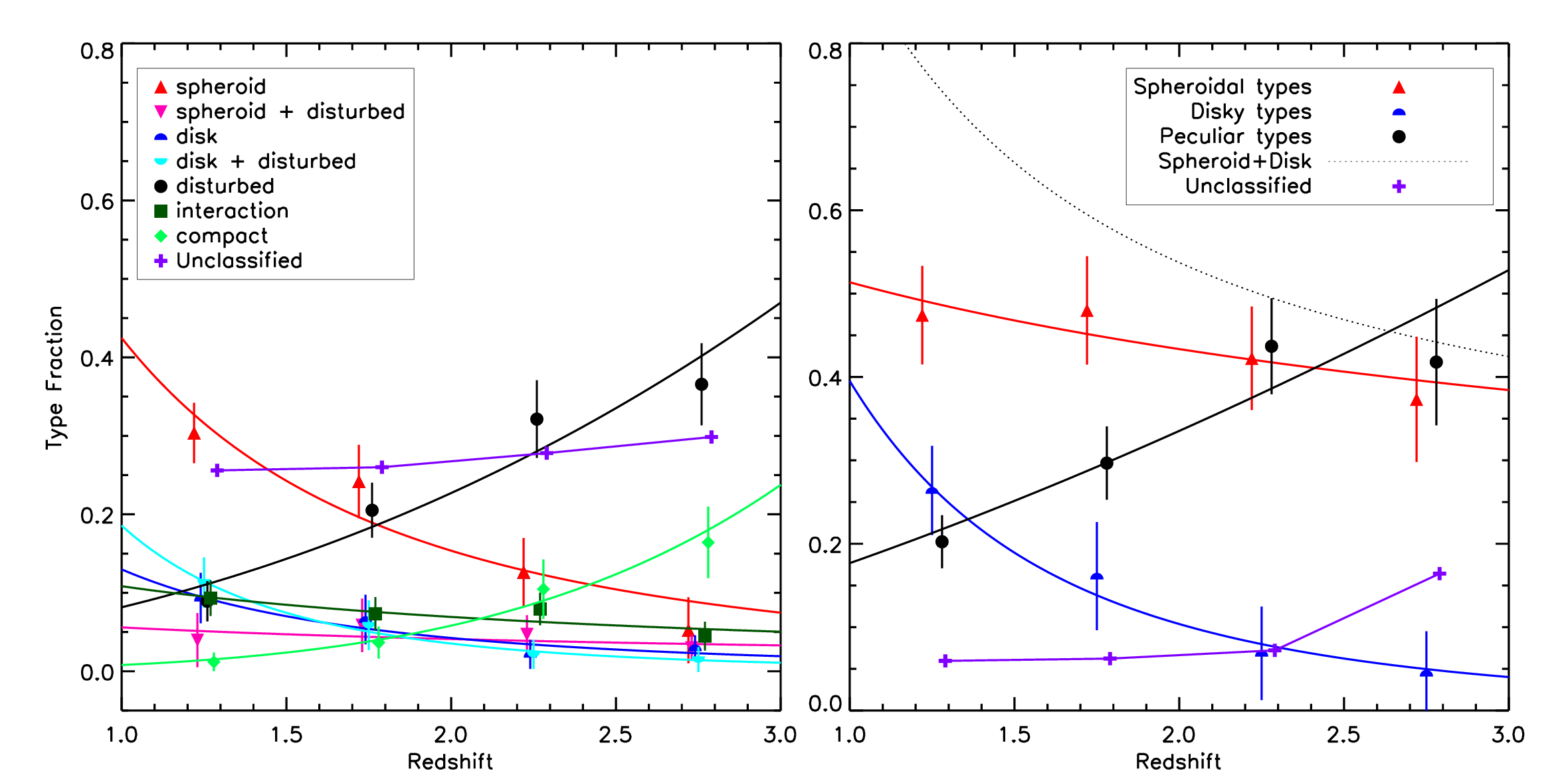}
\caption{The evolution of the fraction of the different galaxy types with redshift before the correction discussed in \ref{sec:typefraccorr}. The left hand panel shows the individual evolution of seven of our galaxy types (Type 0 and Type 8 objects are removed) and the fraction of galaxies which are unclassified. The right hand panel shows the grouped evolution of the galaxy type fraction. The Spheroidal types includes the spheroids, disturbed spheroids and the compact objects. The disky types include the disk and disturbed disk objects. The peculiar types include the disturbed and interacting objects. The errors are a combination of the effect of the stellar mass and redshift errors determined from Monte Carlo simulations and the effects of difference of opinion between classifiers. These are explained fully in Section \ref{sec:errors} The solid lines are power law fits.  The x-axis values are offset by a small amount for clarity.}
\label{galfrac}
\end{figure*}

We also investigate the remaining 302 galaxies that have no classification. We find that generally when we cannot give a galaxy a classification it is due to classifiers disagreeing on the exact galaxy type not the overall galaxy type. For example, the five classifications for a galaxy could be Type 1, Type 1, Type 2, Type 2 and Type 7 which would mean no 3 classifiers agree exactly and hence the galaxy is unclassified. However, there is clearly agreement that overall the galaxy is a spheroid. Therefore, for the right hand panel of Figure \ref{galfrac}, and for subsequent analysis we combine the classifications such that:
\begin{itemize}
\item if three or more classifiers classify the galaxy as a Type 1, Type 2 or Type 7 the galaxy's final classification is spheroid
\item if three or more classifiers classify the galaxy as a Type 3 or Type 4 the galaxy's final classification is disk
\item if three or more classifiers classify the galaxy as a Type 5 or Type 6 the galaxy's final classification is peculiar
\end{itemize}
We find that by this scheme, 1114 (94\%) galaxies receive a classification, this leaves only 74 (6\%) galaxies which we place in the no classification category.

\subsection{Simulations of Galaxy Structure}
\label{sec:cav}
Our definitions of various classifications are based on galaxies in the nearby Universe. This creates a problem because what we would define as one type of galaxy at low redshift may not appear that way at high redshift, either because of the evolution of different galaxy types or because of image quality effects. To try to quantify how this may affect our results we take a sample of Hubble type local galaxies from \citet{Frei96} and a sample of peculiar local galaxies (see \citealt{Cons03}) and artificially redshift them from $z=0$ to $z=1.5$ and $z=2.5$. We do this by reducing the angular size of the galaxy, and then reducing the total galaxy flux by $(1+z)^{4}$ to take into account surface brightness dimming. We then place the galaxy in a simulated background and apply the WFC3 PSF. The final step is to increase the surface brightness of each galaxy by one magnitude because distant galaxies are brighter than nearby galaxies by at least this amount. For a full explanation of this technique see \citet{Cons03}.

\begin{figure*}
\centering
\includegraphics[trim = 13mm 0mm 0mm 6mm, clip,scale=0.45]{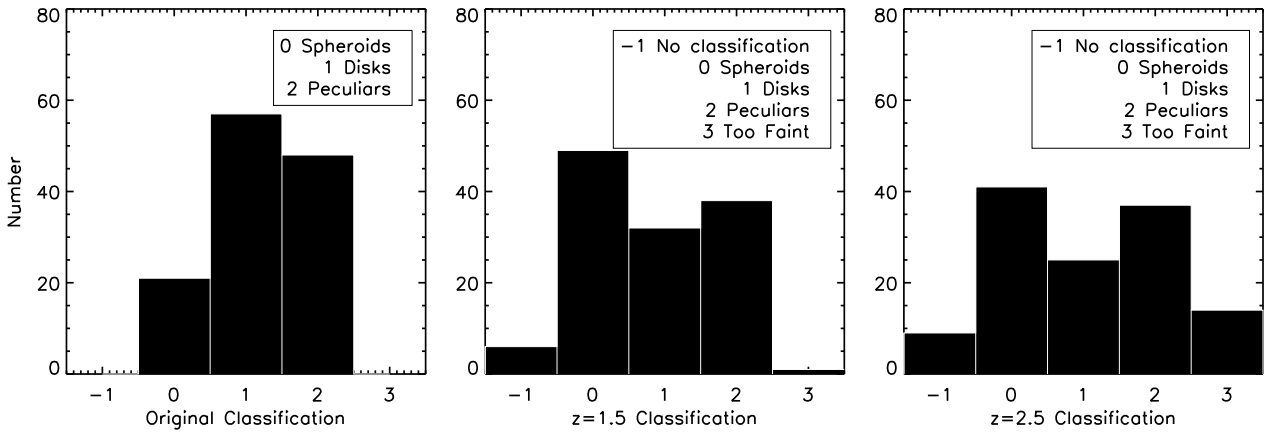}
\caption{The histograms showing the classification of the 82, mostly Hubble type galaxies, from \citet{Frei96} and 44 peculiar galaxies from \citet{Cons03} at $z=0$ (left hand panel) and the classifications of the sample after being artificially redshifted to $z=1.5$ (middle panel) and $z=2.5$ (right hand panel).}
\label{missclass_hist}
\end{figure*}

In the Frei sample there are 82 nearby bright galaxies, imaged in the $R-$band, which have well defined classifications. These objects are chosen to span the Hubble sequence and so are very useful for this task. However, we include the sample of 44 peculiar galaxies, imaged in the $V-$band, to also test these effects on a large disturbed population. Once we have artificially redshifted these galaxies we reclassify them, in this case we use three classifiers and use the two agreeing classifications as our final galaxy type. There are four possible types:
\begin{itemize}
\item Type 0 : Elliptical
\item Type 1 : Disk
\item Type 2 : Peculiar
\item Type 3 : Too faint to classify

\end{itemize}
We find that at $z=2.5$, of the 126 simulated galaxies we classify, only 65 (52\%) are classified as they would be at $z=0$. Of these 65, 17 (26\%) are disk galaxies, 20 (31\%) are spheroids and 28 (43\%) are peculiar galaxies at $z=0$. We also find that of the 126 galaxies there were nine objects (7\%) where no two classifications agreed (one spheroid, five disk and 3 peculiar peculiar galaxies at $z=0$). This leaves 55 galaxies (44\%) where the $z=2.5$ classification does not match the $z=0$ classification. The distribution of misclassifications can be seen in Figure \ref{missclass_hist}. We perform a Kolmogorov–Smirnov (KS) test on these distributions and find probability values which indicate it is very unlikely that the $z=1.5$ and $z=2.5$ distributions are drawn from the distribution of galaxy morphologies at $z=0$. 

These results are slightly better at $z=1.5$, as expected. Of the 126 simulated galaxies, 77 (61\%) are classified as they would be at $z=0$. Of the remaining sample, six (5\%), are not classified due to not enough classifiers agreeing leaving the remaining 43 (34\%) as misclassified. The distribution of misclassifications can also be seen in Figure \ref{missclass_hist}. 

Overall we find that at first glance the problem with misclassification largely effects the disk population, with disks being misclassified as spheroids due to resolution limitations washing out disk structure. We consider the effects of this on our results further in Section \ref{sec:typefraccorr}.

\section{Results}
\label{sec:results}
\subsection{The Morphological Fraction}
\label{sec:typefrac}
We compute the fraction of galaxies of each morphological type at different redshift intervals as shown in Figure \ref{galfrac}. The galaxy fraction in the left hand panel is the number of galaxies in our sample of a given type in a certain redshift range divided by the total number of galaxies in that redshift range. The right hand panel of Figure \ref{galfrac} shows the same quantity but with the various types grouped as described in Section \ref{sec:system}. On both plots the solid lines are simple power law fits of the form $f=f_{0}\times(1+z)^{n}$. The errors on the fractions are discussed in Section \ref{sec:errors}.

In the left hand panels of Figure \ref{galfrac} we see that the fraction of disturbed galaxies shows rapid evolution over the redshift range we are investigating. Approximately 40\% of the galaxies are classified as disturbed galaxies at $z=2.5-3$, this then declines rapidly to a much lower fraction of  $f\sim 0.1$ by $z\sim1$. For the interaction class we see a very small fraction with very little evolution over the whole redshift range.

The disks and disturbed disks both make a negligible contribution in our highest redshift bin, but increase with time until they become comparable to the disturbed population between $z=1-1.5$. The evolution of the pure spheroid class has a form opposite to the disturbed categories. The fraction of spheroids at $z=2.5-3$ is $f\sim0.1$, this increases such that by $z=1.5 - 2$ the spheroid population is the dominant class of galaxy. It is possible that the spheroid point at $z\sim1.75$ is affected by the slightly underdense region at this redshift (this can be seen in Figure \ref{redshiftmass}). However, this point is still within error of the fit. The disturbed spheroid category is almost constant, at roughly a fraction of $f\sim0.1$. The second most dominant type of galaxies between $z=2.5-3$ is the compact population. However, by $z=1-1.5$ the compact population is negligible. As there is no definite size cut between the compact and the spheroidal population there is some ambiguity regarding these two classes. However, the compact population has an average size of $1.21\;kpc$ and the spheroidal population has an average size of $1.76\;kpc$. Therefore, we are selecting the smaller galaxies in the compact population. We note that these are simply visual morphologies and hence may not be the same systems as local galaxies with the same visual morphology.

In the right hand panel of Figure \ref{galfrac}, we combine the subclasses into broader classes and include the galaxies which only have total classifications as discussed in Section \ref{sec:system}. This results in a decrease in the unclassified galaxies (purple crosses) from the left hand panel (uncombined classifications) to the right hand panel (the combined classifications). We find a strong evolution of the peculiar galaxies, similar to that of the individual classifications, from being a large fraction of the total galaxy population at $z>2$ which decreases over the range $z=1-3$.

The total spheroid class in the right hand panel of Figure \ref{galfrac} is a combination of the spheroids, disturbed spheroids and compact populations. It also includes the galaxies which only have a total classification as discussed in Section \ref{sec:system}. Our total spheroid fraction shows there is already a substantial spheroid population at redshift $z\sim3$. When the two disk classes are combined in the right hand panel of Figure \ref{galfrac} we see that the total disk population shows stronger evolution from being almost non existant at $z>2$ to being $\sim 25\%$ of the sample at $z\sim1$. However, the total spheroid population dominates over the total disk population across the whole redshift range. We discuss the emergence of the Hubble type galaxies in Section \ref{sec:hubseq}.

\subsection{The Unclassified Galaxies}
The fraction of unclassified galaxies in the left hand panel of Figure \ref{galfrac} is $\sim25$\% over the whole redshift range. As stated in Section \ref{sec:system} this is due to classifiers disagreeing on specific classification, not overall classification, so this fraction drops considerably in the right hand panel of Figure \ref{galfrac}. We note that of the 302 (25\%) galaxies where three or more classifiers do not agree, 156 (52\%) of these receive an overall classification of spheroid, 46 (15\%) receive an overall classification of disk, 26 (9\%) receive an overall classification of peculiar and 74 (25\%) remain unclassified.

We note that there seems to be a greater difficulty in distinguishing between spheroid, disturbed spheroid and compact. This could be due to the fact that there is some subjectivity in the difference between a spheroidal galaxy and a compact galaxy as we define no definite size cut between these two classes. Furthermore, as spheroids are smooth objects by definition, the disturbances present in these galaxies are often minor and hence this causes disagreement on whether or not they are disturbances worth noting. Further to this, the total spheroid class is comprised of three subclasses, hence disagreement between classifiers can more easily lead to a galaxy obtaining no classification than for the disk and peculiar classes, which are comprised of two subclasses.

\subsection{Corrections to the Type Fractions}
\label{sec:typefraccorr}
\begin{figure*}
\centering
\includegraphics[trim = 11mm 0mm 0mm 6mm, clip,scale=0.46]{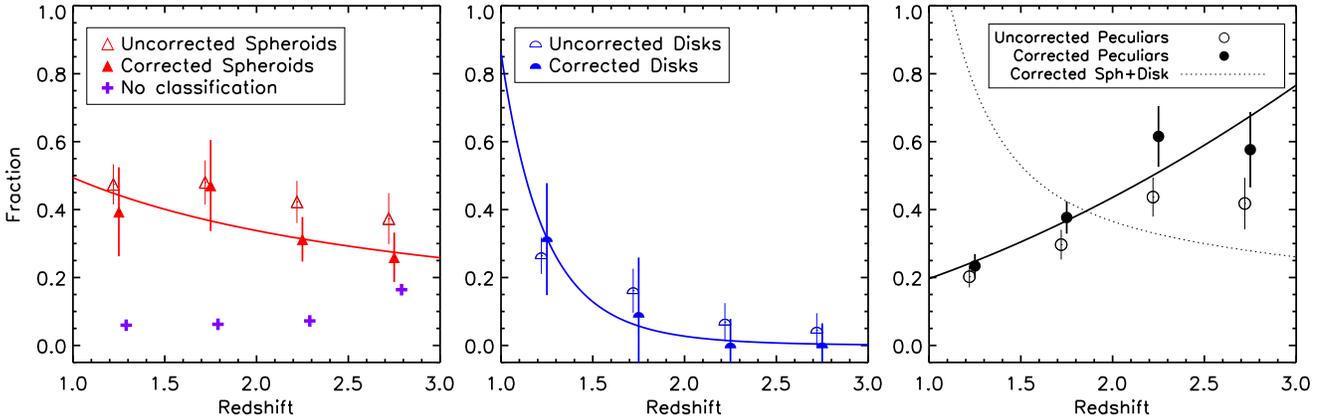}
\caption{The corrected evolution of the fraction of the different galaxy types with redshift grouped as in Figure \ref{galfrac}. The empty points are the original fractions and the filled points are the fractions corrected as discussed in Section \ref{sec:typefraccorr}. The x-axis values are shifted by a small amount for clarity.}
\label{corrfrac}
\end{figure*}

Surface brightness and redshift issues (as discussed in Section \ref{sec:cav}) may have an affect on the morphological fractions computed in this work. At higher redshift it becomes harder to see fainter morphological features (such as a disk) and this can lead to misclassifications. We make a correction for this by taking into account the discrepancies between the classifications for the low redshift galaxy sample (discussed in Section \ref{sec:cav}) and their classifications when artificially redshifted. 

For example, if we examine the spheroid population we can write the true number of spheroids ($N_{S}$) as:

\begin{equation}
N_{S} = N^{'}_{S} - N_{D-S} - N_{P-S} + N_{S-D} + N_{S-P}
\label{eqn:corrfac1}
\end{equation}

\noindent where $N^{'}_{S}$ is the number of objects we classify as spheroids from our observations, $N_{D-S}$ and $N_{P-S}$ are the numbers of true disks and peculiars we have misclassified as spheroids respectively and $N_{S-D}$ and $N_{S-P}$ are the numbers of true spheroids we have misclassified as disk or peculiar respectively

However, we have no information on the number of misclassifications from our observations alone. We therefore make the assumption that
\begin{equation}
\left(\frac{N_{X-Y}}{N_{X}}\right)_{sim}= \left(\frac{N_{X-Y}}{N_{X}}\right)_{obs} = X_{X-Y sim} = X_{X-Y obs}
\label{eqn:corrfac2}
\end{equation}
that is to say, the fraction of galaxies of type X misclassified as type Y ($X_{X-Y}$) is the same in our simulations and our observations. From this we can write
\begin{equation}
N_{X-Y obs} = N_{X obs}\cdot X_{X-Y sim}
\label{eqn:corrfac3}
\end{equation}

\noindent We substitute this into Equation \ref{eqn:corrfac1} and hence write
\begin{equation}
N_{S} = N^{'}_{S}\rm -\rm N_{D}\cdot X_{D-S}\rm - \rm N_{P}\cdot X_{P-S} \rm+ \rm N_{S}\cdot(X_{S-D} + X_{S-P}).
\label{eqn:corrfac4}
\end{equation}
This can be rearranged to
\begin{equation}
N_{S} = \frac{N^{'}_{S}\rm -\rm N_{D}\cdot X_{D-S}\rm - \rm N_{P}\cdot X_{P-S}}{1-(X_{S-D} + X_{S-P})}
\label{eqn:corrfacsph}
\end{equation}
where $X_{D-S}$ is the fraction of disk galaxies misclassified as spheroids in our simulations, $X_{P-S}$ is the fraction of peculiar galaxies misclassified as spheroids in our simulations, $X_{S-D}$ is the fraction of spheroids misclassified as disks in our simulations and $X_{S-P}$ is the fraction of spheroids misclassified as peculiars in our simulations. $N_{S}$, $N_{D}$ and $N_{P}$ are the true number of spheroids, disks and peculiars respectively.

We can construct similar equations for both the disk and peculiar population using the same method. We can then write that
\begin{equation}
N_{D} = \frac{N^{'}_{D}\rm -\rm N_{S}\cdot X_{S-D}\rm - \rm N_{P}\cdot X_{P-D}}{1-(X_{D-S} + X_{D-P})}
\label{eqn:corrfacdisk}
\end{equation}
and also that
\begin{equation}
N_{P} = \frac{N^{'}_{P}\rm -\rm N_{S}\cdot X_{S-P}\rm - \rm N_{D}\cdot X_{D-P}}{1-(X_{P-S} + X_{P-D})}
\label{eqn:corrfacpec}
\end{equation}
where $N^{'}_{D}$ and $N^{'}_{P}$ are the observed numbers of disk and peculiar galaxies, $X_{P-D}$ is the fraction of peculiar galaxies misclassified as disks in our simulations and $X_{D-P}$ is the fraction of disk galaxies misclassified as peculiar in our simulation. 

\begin{figure*}
\centering
\includegraphics[trim = 10mm 0mm 0mm 8mm,clip,scale=0.45]{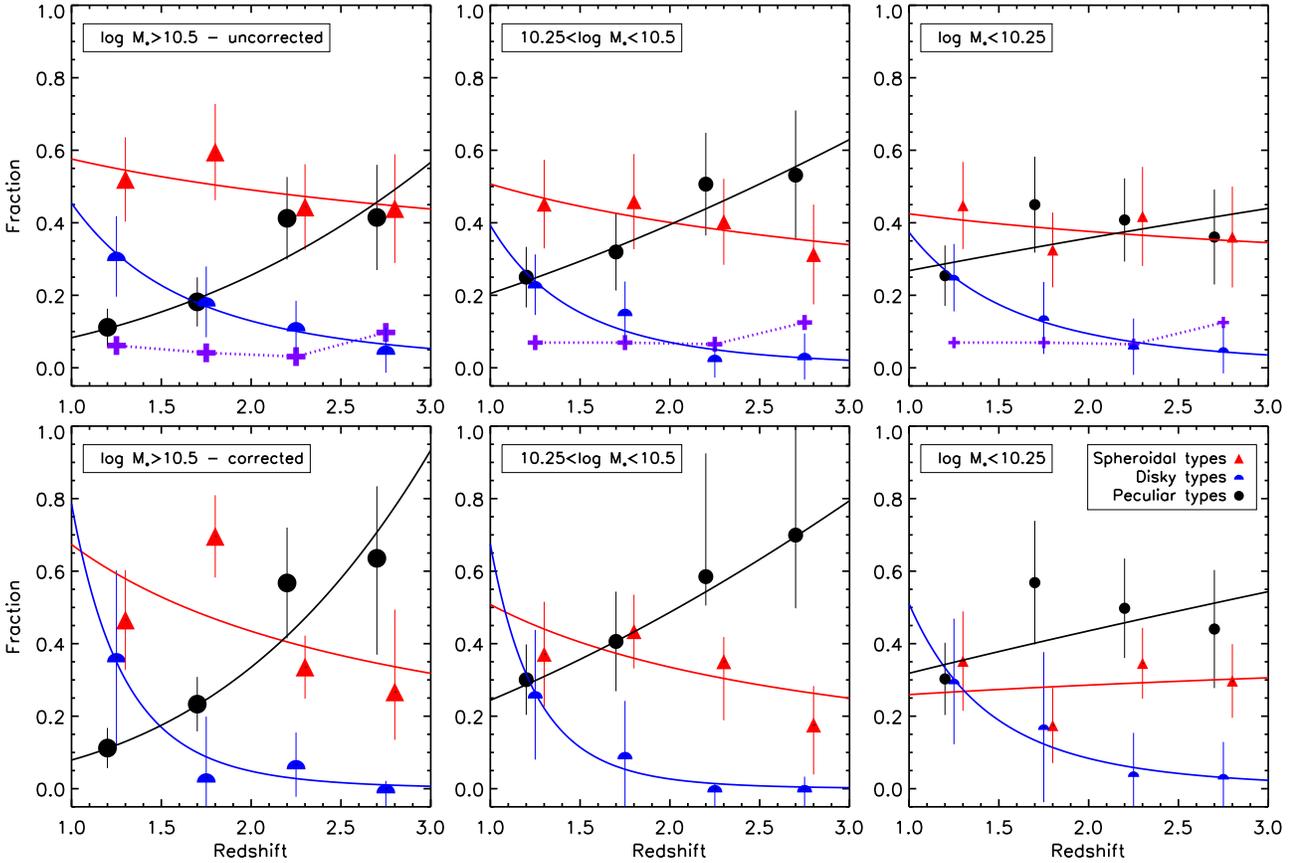}
\caption{The galaxy type fraction split by mass as a function of redshift. The size of the points increases with stellar mass. The lines are simple power law fits. The x-axis values are offset by a small amount for clarity.}
\label{typemassfrac}
\end{figure*} 

Therefore, we can solve this set of equations simultaneously, and find the corrected numbers of spheroid, disk and peculiar galaxies. For the $z=1-2$ redshift range we take the simulated fractions from the \citet{Frei96} and \citet{Cons03} galaxies artificially redshifted to $z=1.5$. For the $z=2-3$ redshift range we artificially redshift the \citet{Frei96} and \citet{Cons03} galaxies to $z=2.5$. We then use these in Equations \ref{eqn:corrfacsph}, \ref{eqn:corrfacdisk} and \ref{eqn:corrfacpec} to calculate the corrected fraction which we normalise so that no galaxy fraction can be less than zero. These corrected fractions are plotted in Figure \ref{corrfrac}.

For the spheroids, the correction reduces the fraction at all redshifts. This is not entirely surprising as resolution problems cause galaxies to be smoothed out and loose structure, hence appearing spheroidal. We find a large increase in the peculiar fraction after correction at $z>2$. This is interesting because it could be argued that it is harder to confuse disturbed structure with smooth structure or a disk, therefore, these would be the easiest class of galaxy to classify. However, from the decrease in the spheroid fraction and the increase in peculiar fraction, we infer that the disturbed structure is faint in these galaxies, hence they appear too smooth and are misclassified.

The most striking feature about Figure \ref{corrfrac} is that at $z>2$ there is an apparent lack of any disk galaxies. This would seem to be in disagreement with \citet{Bruc12} who conduct analysis of the S\'{e}rsic indices  (we directly compare with S\'{e}rsic indices in Section \ref{sec:nodisks}) and detailed bulge disk decompositions of the most massive galaxies ($M_{*}\ge10^{11}M_{\odot}$) used in this work. They find that the $z>2$ Universe is dominated by galaxies whose bulge to disk ratios suggest they are disky.

We note that the correction at $z=2-3$ is more severe than at $z=1-2$. It is not surprising that at higher redshift there are more image quality problems. It is encouraging to note that the correction makes little difference to our morphological fractions, and this gives us confidence that by using this data we have produced reliable results. We discuss these results further in Section \ref{sec:hubseq}. One caveat of our correction is that our local galaxies may be intrinsically brighter than our high redshift sample and hence easier to classify. However, we test how our corrections are affected by matching the magnitude of our simulated galaxies to that of the average magnitude of the lowest stellar mass galaxies ($M_{*}<10^{10.25}M_{\odot}$) in our sample in Section \ref{sec:galmass}. We find that this has little affect on our corrected fractions.

\subsection{Errors on the Morphological Fraction}
\label{sec:errors}
To quantify the errors on the morphological fraction we include errors from the photometric redshifts, masses, number statistics and disagreement between classifiers. To take into account the number statistics involved in our analysis we calculate simple Poissonian errors for each redshift bin. For the photometric redshift and stellar mass errors we use a Monte Carlo approach which randomly varies the measured redshift or stellar mass within the error. We then recalculate the morphological fraction based on the new simulated photometric redshifts and masses. We repeat this process 1000 times and then take the standard deviation of the simulated fractions as the final error.

As our stellar masses are calculated from the mode of a distribution of fitted templates as discussed in Section \ref{sec:zandm}, there is a possible error that arises from binning the stellar masses. We therefore also include an additional Monte Carlo variation to the stellar masses of between plus or minus the bin size used (0.05 dex). We find that this additional variation has only a small effect on the stellar masses and hence the morphological fractions.

We include the uncertainty due to the disagreement between individual classifiers by comparing the fractions as though they were calculated using the classifications of each individual classifier. We include this in the error by taking the standard error on the mean ($\frac{\sigma}{\sqrt(N)}$) of each fraction for each classifier. We then add this in quadrature to our errors from the redshift and stellar mass Monte Carlo analysis and Poissonian error. We take this quantity as our total error.

For the corrected fractions we also include the error due to disagreement between classifiers when reclassifying the simulated galaxies. We calculate the corrected fractions for each individual classifier's results and again we take the standard error on the mean of these fractions. We also apply the correction to $f\pm\Delta f$ and obtain a upper and lower corrected fraction. We use the differences between these two extremes and the measured corrected fraction as part of the error. We add the appropriate errors in quadrature and take this quantity as our total error.

\subsection{Galaxy Fractions and Stellar Mass}
\label{sec:galmass}
Thanks to the large sample size we can split our sample by stellar mass and explore visual morphology as a function of both redshift and stellar mass. Figure \ref{typemassfrac} shows the galaxy type fractions split into stellar mass bins of $M_{*}\ge10^{10.5}M_{\odot}$, $10^{10.5} M_{\odot} > M_{*} \ge 10^{10.25} M_{\odot}$, and $M_{*}<10^{10.25}M_{\odot}$. The panels increase in stellar mass from left to right and the top and bottom panels are the fractions uncorrected and corrected respectively. The symbols and colours have the same meaning as in the right hand panel of Figure \ref{galfrac}, and the errors are calculated as in Section \ref{sec:errors}.

For the $M_{*}<10^{10.25}M_{\odot}$ bin, we apply a different correction than for previous plots. We use artificially redshifted galaxies as before, however, we also artificially dim the galaxies to the average $H_{160}-$band magnitude of the galaxies in this bin ($H$ mag = 22.65 at $z<2$ and $H$ mag = 23.68 at $z>2$). In this way, we account for the fact that the misclassifications due to image quality will depend on galaxy brightness (i.e. stellar mass).

After correction we find that the emergence of the Hubble type galaxies, the disks and spheroids, depends on stellar mass. For the most massive galaxies ($M_{*}\ge10^{10.5}M_{\odot}$) we find that the fraction of spheroids is comparable to that of the peculiars even at the highest redshift. These high stellar mass spheroids then begin to dominate between $z=2-2.5$. We find this transition increases in redshift as we go to the lower mass bins such that the transition is between $z=1-1.5$ in the lowest mass bin. We see a similar trend for the disk population where we find that they dominate over the peculiar galaxies between $z=1-1.5$ for the $M_{*}\ge10^{10.5}M_{\odot}$ galaxies but the fraction of disks is only comparable to the peculiars, in that redshift range, for the other two mass bins.

We note that we are slightly incomplete to red objects in the $z\sim3$ fraction in the lowest mass bin. To test if this affects our results we repeat our analysis with galaxies which have $M_{*}\ge10^{10.25}M_{\odot}$ and still find the same result.

\subsection{The Evolution of the Number Density of Galaxy Types}
\label{sec:numden}
\begin{figure}
\centering
\includegraphics[trim = 0mm 0mm 0mm 6mm, clip,scale=0.5]{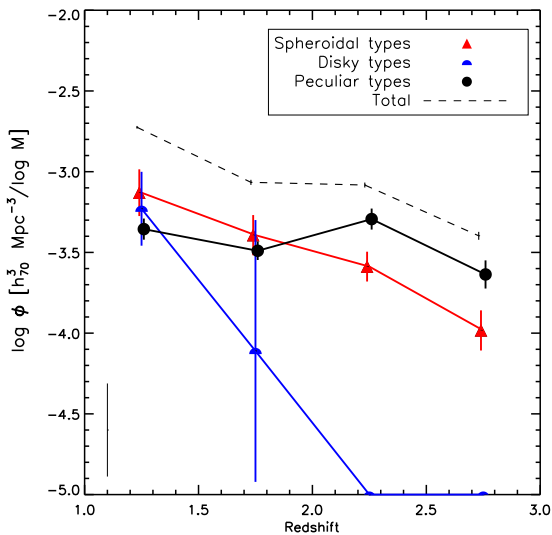}
\caption{The evolution of the total number density of galaxies between $z=1-3$ (dashed black line). Also plotted is the evolution of the number density of each galaxy type. The symbols and colours are as in Figure \ref{galfrac}. The x-axis values are shifted by a small amount for clarity. We have added $1\times 10^{-5}$ to the $z>2$ number densities of the disk galaxies to avoid taking the log of zero. The error bar in the bottom left corner shows the typical error from cosmic variance.}
\label{numberden}
\end{figure}
\begin{figure}
\centering
\includegraphics[trim = 0mm 0mm 0mm 6mm, clip,scale=0.5]{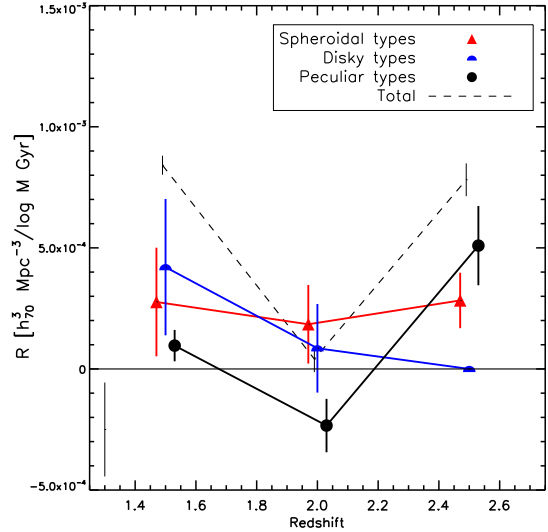}
\caption{The rate of growth of galaxies between $z=1-3$.  The dashed black line is the rate of growth of all of galaxies in our sample. Also plotted is the rate of growth of galaxies of each type. The symbols and colours are as in Figure \ref{galfrac}. The x-axis values are shifted by a small amount for clarity. Negative values indicate a decline in the number density. The error bar in the bottom left corner shows the typical error from cosmic variance.}
\label{rates}
\end{figure}
Using our fractions of each galaxy type we can investigate the evolution of the total number density of galaxies in our sample, as well as the evolution of the number density of each visual galaxy population. This is shown in Figure \ref{numberden}, where the evolution of the total galaxy population is shown by the dashed black line, and the evolution of the number density as a function of morphological type is shown by the red, blue and black symbols (colours are the same as those in the right hand panel of Figure \ref{galfrac}). The errors on the total number density are from our Monte Carlo analysis as discussed in Section \ref{sec:errors}. We show the typical error from cosmic variance, computed as in \citet{Most11}, in the bottom left corner of Figure \ref{numberden}. The errors on the number densities for each fraction arise from error propagation on the total number density and the fraction of each type.

We find that, despite the rapid evolution of the fraction of peculiar galaxies, the number density of this type evolves very little in our redshift range. We see stronger evolution in the disk and spheroidal types and it is the emergence of these Hubble type galaxies which is driving the evolution in the total number density.

We can investigate the rate of massive galaxy formation, both total and for individual classes of galaxies, by looking at how the number density of galaxies changes with redshift. We calculate the rate of growth of our galaxies as:
\begin{equation}
R = \frac{\delta\phi(z)}{\delta t}
\end{equation}
 where $\delta\phi(z)$ is the difference between the number density of two redshift bins, and $\delta t$ is the time in the redshift range being considered. The rate of growth of the total galaxy population and each classification of galaxy can be seen in Figure \ref{rates}. The total rate is the black dashed line and the black, red and blue solid lines are the fractional rates for the different classifications of galaxies. Colours and points have the same meaning as in the right hand panel of Figure \ref{galfrac}. The errors are propagated from the errors on the number densities in Figure \ref{numberden}. We show the typical error from cosmic variance in the bottom left corner of Figure \ref{rates}.

We are limited to three data points due to our redshift bins but we find tentative evidence for a constant rate of growth for all galaxies. However, we show a difference in the rate of growth for different types of galaxies. We note that the errors are dominated by cosmic variance in both Figure \ref{numberden} and Figure \ref{rates}. This implies that, in the redshift range we probe, our results may differ for a global study. However, in Mortlock et al. (in preparation) we construct the full galaxy stellar mass function for the the CANDELS UDS and find good agreement with previous measures of the galaxy stellar mass function. Therefore, our measures of the total number density and rate of growth agree well with the literature.

\section{Analysis}
\label{sec:diss}
\subsection{Quantifying the Emergence of the Hubble Sequence}
\label{sec:hubseq}
In this study, we show the decrease in the fraction of peculiar galaxies and the increase of the `normal' disk and spheroid populations between redshift $1<z<3$ (Figure \ref{corrfrac} and Figure \ref{typemassfrac}). We define $z_{trans}$ as the redshift where the fraction of peculiar galaxies is equal to the total fraction of spheroid and disk galaxies, i.e. where $f_{pec}(z)=f_{sph}(z)+f_{disk}(z)$.  We find the redshift where the total Hubble population becomes dominant, for all galaxies with $M_{*}\ge10^{10}M_{\odot}$, as $z_{trans}=1.86\pm0.62$. This allows us to quantify the emergence of the Hubble sequence whereby the high redshift disturbed population settle down into the galaxies we see in the local Universe.

It is possible our classification criteria are affecting the value of $z_{trans}$. We test this in several ways:\\
1) We recalculate the morphology fraction, and hence $z_{trans}$, with the criteria that 4 or 5 classifiers need to agree for the galaxy to be given a classification.\\
2) We calculate $z_{trans}$ for each individual classifier.\\
3) We remove the interaction class and use the classification of the central galaxy only.\\
4) We include disturbed spheroids and disturbed disks in the total peculiar population.
We find that $z_{trans}$ differs significantly for one specific classifier. However, this is not evidence that $z_{trans}$ is wrong, simply evidence that it is incorrect to base visual classifications on one classifier's results. Using criterion 4 we find a low value of $z_{trans}$, however this is expected because of the structure of our classification scheme. The disturbed spheroids and disturbed disks are Hubble type galaxies with small disturbances by construction of our classification scheme. Therefore, we are including Hubble type galaxies in the disturbed population and making the disturbed fraction too high. We also test the effect of size evolution on our simulated galaxy classifications, and how this affect $z_{trans}$. We know galaxies at high redshift are smaller than their local counterparts, and that this is also different for galaxies of different morphology. We perform our simulations again on the \citet{Frei96} and \citet{Cons03} local galaxies but this time evolve the sizes of these galaxies according to \citet{Buit08} and find that the distribution of sizes of the simulated galaxies match well our sample. We reclassify these galaxies at $z=2.5$ (where the affects of size evolution would be the largest) and find the fractions, when corrected including size evolution, are comparable to those in Figure \ref{corrfrac}. Therefore, this has a negligible affect on the value of $z_{trans}$. 

\subsection{Comparison to Previous Studies}
\label{sec:compprevwork}
It is useful to compare our work to studies that have investigated visual morphology of galaxies with similar stellar mass limits and redshift ranges. We find that that the dominant peculiar population at high redshift, and that the cross over to a Universe dominated by Hubble type galaxies, is in good agreement with previous studies (e.g. \citealt{Driv98}, \citealt{Cons05}, \citealt{Papo05}, \citealt{Came11}, \citealt{Szom11}). The large peculiar population at high redshift is not unexpected as this is when galaxies are in formation and are less dynamically settled (e.g. \citealt{Cons08}).

We note also, that this transition of the visual morphologies of galaxies in the redshift range $1<z<3$ is supported by the proposed formation scenario in the recent work of \citet{Driv13}. By examining the star-formation history of spheroid and disk galaxies, they infer that the Universe above redshift of $z\sim2$ is dominated by the merger driven formation of spheroids. After redshifts of $z\sim2$, the dominant formation mechanism switches to cold gas accretion and hence, the formation of disk galaxies dominates. This agrees well with our already substantial spheroid presence and our lack of disk galaxies at $z>2$. These results are further supported by \citet{Cons13} who show that cold gas accretions plays a vital role in the the formation of galaxies in the epoch studied here.

Overall we find good agreement with several previous studies and this work expands on these by including a much larger number of galaxies with deeper imaging and hence more robust visual morphologies. This has allowed us to quantify the emergence of the Hubble sequence in detail in Section \ref{sec:hubseq}.

\subsection{The Dependence on Stellar Mass}
Thanks to our large galaxy sample, we have the number statistics to investigate the evolution of the visual morphology fraction as a function of stellar mass. As mentioned previously there are several studies whose results, at first glance, differ from ours. Both \citet{Cons11} and \citet{Buit13} find that even at $z\sim3$ the Hubble types are the dominant galaxy populations. However, these studies focus on the most massive galaxies. That is, galaxies with $M_{*}\ge10^{11}M_{\odot}$, and hence it makes more sense to compare these works to the left hand panel of Figure \ref{typemassfrac}.

We find in Figure that \ref{typemassfrac} the emergence of the Hubble sequence depends on stellar mass. The highest mass galaxies in the Universe are dominated by Hubble type galaxies at an earlier epoch than the lower mass galaxies. We calculate $z_{trans}$ for each mass bin and find the transition redshift for the high, intermediate and low mass bins to be $z_{trans}=2.22\pm0.82, z_{trans}=1.75\pm 0.73$ and $z_{trans}=1.73\pm0.57$ respectively.

Therefore, our results agree with \citet{Cons11} and \citet{Buit13} who find a large fraction of Hubble morphologies for massive galaxies. Thus we conclude that the most massive galaxies become morphologically settled first. We know from the evolution of the galaxy stellar mass function that the most massive galaxies have assembled most of their stellar mass by $z\sim2.5 - 3$ (e.g. \citealt{Mort11}). We also know that massive galaxies complete their major episodes of star-formation before their lower mass counter parts (\citealt{Baue05}, \citealt{Feul05}, \citealt{Bund06}, and \citealt{Verg08}).

Complementary to this, it has been shown that mergers are an important mechanism in galaxy formation. Recent studies on the evolution of the merger fraction of galaxies (e.g. \citealt{Bluc09}, \citealt{Bluc12}, \citealt{Man12}) show that the most massive galaxies are undergoing most of their mergers at $z\sim 2-3$. Furthermore, \citet{Bluc09} find that the merger fraction for massive galaxies turns over at higher redshift than for low mass galaxies. Therefore, we explain the result of the dependence of the emergence of the Hubble sequence on stellar mass, as morphological downsizing driven by mergers, whereby the most massive galaxies are settled first.

\subsection{Why Are There So Few Disk Galaxies at $z>2$?}
\label{sec:nodisks}
\begin{figure*}
\centering
\includegraphics[trim = 0mm 0mm 0mm 6mm, clip,scale=0.35]{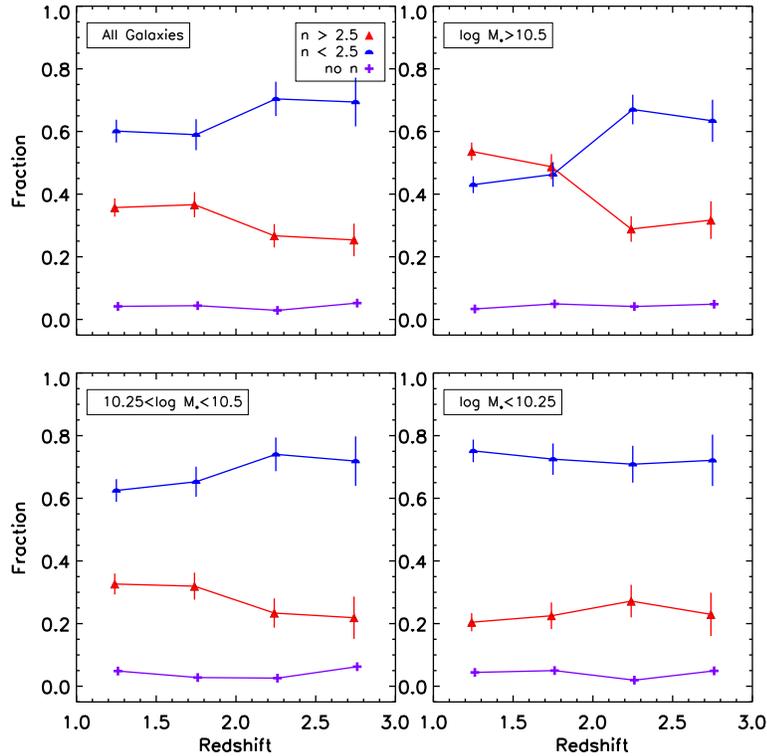}
\caption{The evolution of the total fraction of galaxy types defined using S\'{e}rsic indices and also split by mass. `Spheroid like' galaxies (red triangles) have S\'{e}rsic index greater than 2.5 and `disk like' galaxies (blue squares) have S\'{e}rsic indices less than 2.5. The top left panel is the total fraction for all galaxies with $M_{*}\ge10^{10}M_{\odot}$. The remaining three panels are split by mass according to the legend. The purple crosses are galaxies for which no S\'{e}rsic index could be measured.}
\label{sersic}
\end{figure*}
In this work we find that the visual disk population is nonexistent at $z>2$. We compare this result to the S\'{e}rsic index evolution in Figure \ref{sersic} where we split our sample into S\'{e}rsic spheroids ($n>2.5$) and S\'{e}rsic disks ($n<2.5$) using the S\'{e}rsic indices from \citet{Vand12}. The errors are from a Monte Carlo analysis where we alter the stellar masses, photometric redshifts and S\'{e}rsic indices between their measured errors, then recalculate the fraction. We do this 1000 times and take the error as the standard deviation.

We find that for the massive galaxies (top right panel) we agree with the results of \citet{Bruc12} who find that at high redshift, galaxies with low S\'{e}rsic indices dominate until around $z\sim2$. When we look at the total population (top left) and the lower mass galaxies (bottom panels) we do not find this cross over, and the low S\'{e}rsic index populations are dominant at all redshifts. Further to this, other studies have shown galaxies have disk-like structure at high redshift based on light profiles (e.g. \citealt{Yuma11}, \citealt{Vand11}, \citealt{Chan13}, \citealt{Buit13}, \citealt{Pate13}).

Within our sample, 154 (53\%) Of the $z>2$ low-S\'{e}rsic index galaxies, are visually classified as peculiar. We expect peculiar galaxies to have low S\'{e}rsic indices as they are often elongated. This suggests that at these redshifts it is more accurate to say that S\'{e}rsic index is tracing how extended a galaxy is, and not as an indicator of the Hubble type morphology.

Our lack of galaxies with visual disk morphologies also suggests the Hubble tuning fork is not suited to galaxies at high redshift. There are many possible reasons why, at $z>2$, there are no galaxies which fit the Hubble tuning fork visual disk classification. One scenario is that these disks are already formed but are being disturbed by one or more of several processes which may play an important role in galaxy evolution at high redshift. We know from simulations (e.g. \citealt{Deke09}, \citealt{Ceve10}) that rotating disks can undergo violent disk instabilities which result in clumpy disrupted morphologies. This is also supported by observations of rotating galaxies at $z>2$ (e.g. \citealt{Elme07}, \citealt{Genz08}, \citealt{Genz11}). 

Further to this, we know from studies of minor mergers at high redshift (e.g. \citealt{Lotz11}, \citealt{Bluc12}) that galaxies can undergo many such mergers in their lifetime. It could be that this process is causing the disruption of disks at high redshift, making them appear distinct from the $z=0$ disk population. Feedback is a further mechanism by which a galaxy can be disrupted. It is unclear which (if any) of these possible scenarios is the main cause of the disturbed disk population at $z>2$.

Alternatively, these disks may be in the early stages of their formation at these epochs and hence will contain clumps and structural peculiarities. The fainter disk of a galaxy will be dominated by the bright clumpy features, therefore the galaxy will appear visually disturbed. This would lead to a galaxy being classified as peculiar and hence contributing to the peculiar fraction. This agrees with previous studies such as \citet{Krie09} and \citet{Law12} who find that the star forming population at high redshift is irregular and also argue that classic star forming disks do not exist at high redshift. Furthermore, \citet{Wuyt12} look at the stellar mass and stellar light profiles of galaxies within CANDELS and show that they often show clumpy features. However, these clumps are often found in the light profiles of these galaxies and not in the structural distribution of the stellar mass. This further highlights the difficulties of infering properties from the light profiles of these clumpy galaxies.

It is unclear which (if any) of these possible scenarios is the main cause of the lack of a settled disk population at $z>2$. Therefore, we interpret our visual morphology results, not as an indication that truly no disks exist at high redshift, but that disks which have the same visual morphologies as the classic disks we see in the local Universe are rare at $z>2$. If we are to fully understand the evolution of disk galaxies it is important to understand these processes. This suggest that to understand the formation of disk galaxies, and to separate them from merging systems, we need to study the redshift range $z=2-3$ in detail, particularly the kinematics of these objects.

\subsection{Visual Morphology and Star Formation}
\label{sec:sfr}
\begin{figure*}
\centering
\includegraphics[trim = 12mm 0mm 0mm 6mm, clip,scale=0.45]{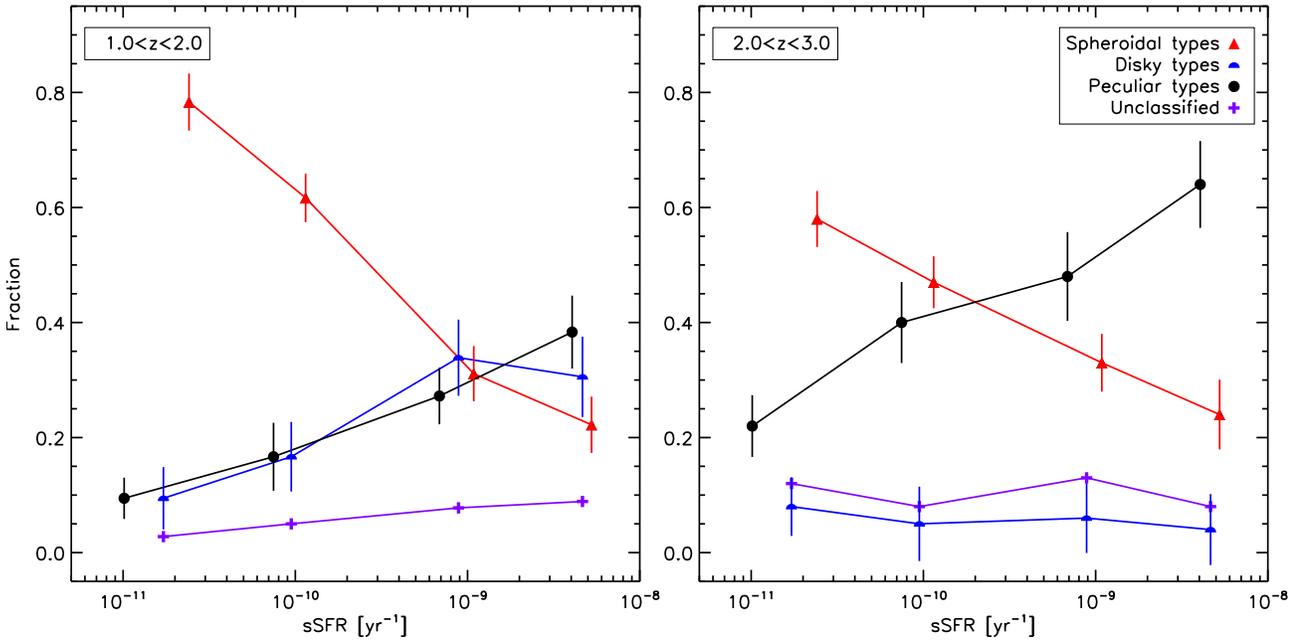}
\caption{The evolution of the fraction of each morphological classification of galaxies as a function of sSFR for our sample of $M_{*}\ge10^{10}M_{\odot}$ galaxies. The lowest sSFR bin contains the least star forming galaxies and the highest sSFR bin contains the most star forming galaxies. The points are coloured by type as in the right hand panel of Figure \ref{galfrac}. The x-axis values are shifted by a small amount for clarity.}
\label{sersicsfr}
\end{figure*}

In the local Universe we know there is a link between visual morphology and star formation rate. To investigate if this is also the case at $z>1$ we plot the fraction of each type of visual classification, as in the right hand panel of Figure \ref{galfrac}, as a function of specific star formation rate ($sSFR=SFR/M_{*}\: [yr^{-1}]$) in Figure \ref{sersicsfr}. We split the sample into four bins containing approximately equal numbers of galaxies.

The errors include a Monte Carlo analysis and the error from the visual classifications. For the Monte Carlo analysis we randomly vary redshift, sSFR and stellar mass between their errors and recalculate the fraction as a function of sSFR and redshift each time. We do this 1000 times and take the error as the standard deviation of these. For the error from the visual classifications we recalculate the fraction as a function of star formation rate and redshift for each classifiers results and take the standard error on the mean as the error. We add these two components in quadrature to obtain the total error.

We find that the least star forming galaxies are likely to be part of the spheroid population. Also, we find that, generally, the most star forming galaxies (i.e. in the highest star formation bins) are most likely to be peculiar at $z>2$. At $z<2$ galaxies in the highest star forming bin are roughly half spheroidal and half extended (the sum of disk and peculiar visual types). This indicates a link between the star-formation rate and the visual morphology of a galaxy.

\subsection{CAS Morphology}
\begin{figure}
\centering
\includegraphics[trim = 0mm 0mm 0mm 6mm, clip,scale=0.4]{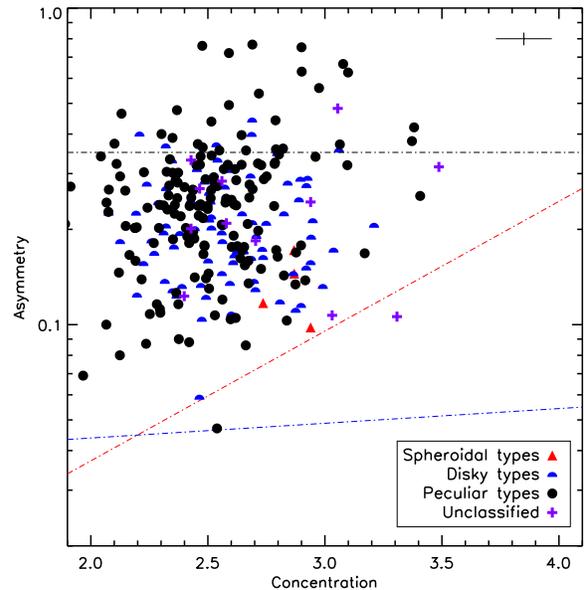}
\caption{The classification of the galaxies in this sample using the concentration and asymmetry values and the cuts derived for local galaxies from \citealt{Bers00}. The blue line is the separation between early and intermediate types and the red line is the separation between intermediate and late type. Galaxies above the black line ($A>0.35$) are considered to be mergers. The points are coloured by visual classification as in the right hand panel of Figure \ref{galfrac}}
\label{cas}
\end{figure}

Another method of looking at galaxy morphology is to use the CAS parameters. \citet{Bers00} classify a sample of low redshift galaxies using their concentration and asymmetry indices and define regions populated by different galaxy populations. Figure \ref{cas} shows our sample of galaxies split into late (above red line), intermediate (between red and blue line) and early (below blue line) types using their relations. We plot the galaxies with effective radii, as measured by CAS, greater than $\sim$0.4 arcsecs. This leaves only 269 galaxies from our sample. However, we are confident we are looking at galaxies which are resolved enough to compare to local galaxies (\citealt{Cons00}), and hence the structure of the galaxy is clear.

We find that almost all of our galaxies are classified as late (disk) types or mergers according to this method, suggesting there is very little morphological diversity in the redshift range $z=1-3$ using CAS. Even if we include the galaxies which have less reliable CAS measurements we find that only 12\% are not CAS late types. This is in disagreement with our visual morphologies and suggests an ever higher disk-like fraction than implied by their S\'{e}rsic indices. 

However, these cuts are not expected to define well morphologies of galaxies at these redshifts as they were calibrated on low redshift galaxies. In this sense this can be taken as further evidence that high redshift galaxies are structurally distinct from their low redshift counterparts. This is in agreement with \citet{Bluc12} who show that galaxy asymmetry rises with redshift, while concentration decreases with redshift. Hence, galaxies shift out of the regions defined in \citet{Cons03} at progressively higher redshifts.

\subsection{Rest frame U-B Colour}
\begin{figure*}
\centering
\includegraphics[trim = 0mm 0mm 0mm 6mm, clip,scale=0.45]{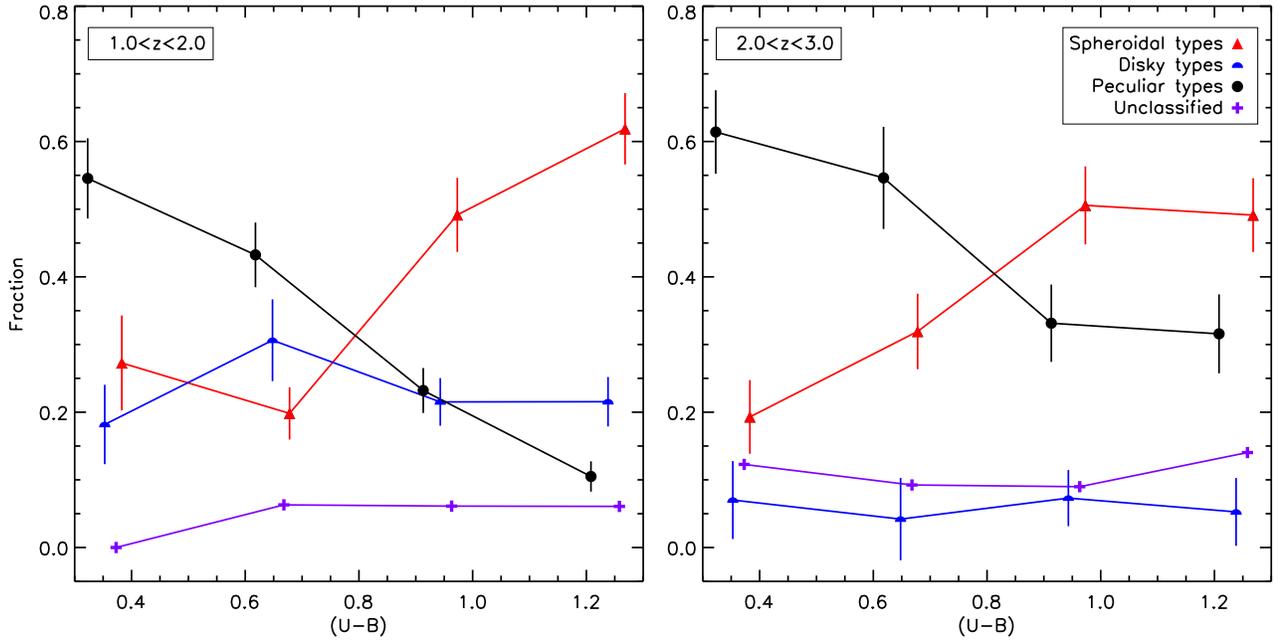}
\caption{The evolution of our morphological fractions as a function of rest frame $U-B$ colour, uncorrected for dust, for our sample of $M_{*}\ge10^{10}M_{\odot}$ galaxies. The bin with the lowest $U-B$ value contains the bluest galaxies and the bin with the highest $U-B$ value contains the reddest galaxies. The points are coloured as in the right hand panel of Figure \ref{galfrac}. The x-axis values are shifted by a small amount for clarity.}
\label{ubfrac}
\end{figure*}

\begin{figure*}
\centering
\includegraphics[trim = 0mm 0mm 0mm 6mm, clip,scale=0.45]{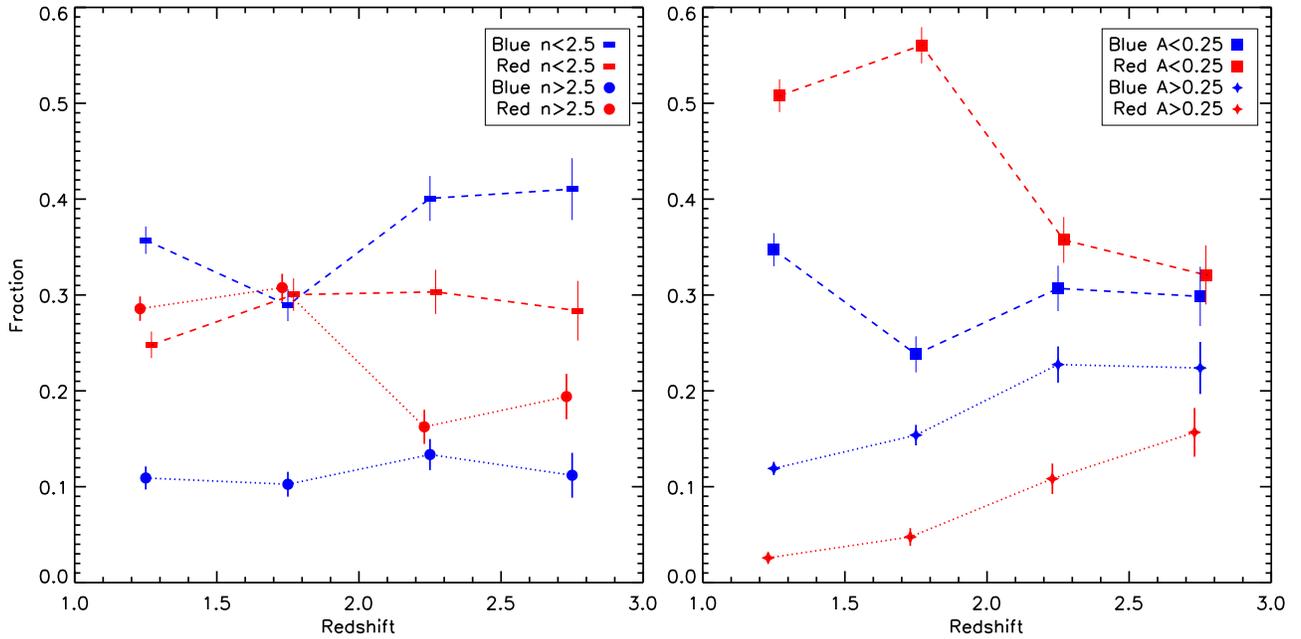}
\caption{Left hand panel: The fraction of galaxies in our sample split by $U-B$ colour and S\'{e}rsic index as a function of redshift for our sample of $M_{*}\ge10^{10}M_{\odot}$ galaxies. The blue rectangles are the blue $n<2.5$ fraction, the red rectangles are the red $n<2.5$, the blue circles are the blue $n>2.5$ fraction and the red circles are the red $n>2.5$ fraction. Right hand panel: The fraction of galaxies split by $U-B$ colour and asymmetry as a function of redshift. The blue stars are the blue $A<0.25$ fraction, the red stars are the red $A<0.25$, the blue squares are the blue $A>0.25$ fraction and the red squares are the red $A>0.25$ fraction. The x-axis values are shifted by a small amount for clarity.}
\label{ubparams}
\end{figure*}

We explore possible links between visual morphology and rest frame $U-B$ colour in Figure \ref{ubfrac}. We plot the morphological fraction of galaxies (as in Figure \ref{galfrac}) as a function of rest frame $U-B$. In Figure \ref{ubparams} we divide the galaxies into red and blue using the equation from \citet{Will06}, modified for the AB magnitude system. The equation is written as
\begin{equation}
-0.032(M_{B}-\Delta M_{B}+21.52)+1.284-0.25+\Delta(U-B)
\label{eq:colcut2}
\end{equation}
\noindent where $M_{B}$ is the restframe $B-$band magnitude of the galaxy and $\Delta M_{B}$ and $\Delta(U-B)$ are the corrections for redshift evolution from \citet{Vand01}. We then apply the cut so that if the restframe $(U-B)$ colour is greater than Equation \ref{eq:colcut2} the galaxy is red, and if $(U-B)$ is less than Equation \ref{eq:colcut2} the galaxy is blue.

In the left hand panel of Figure \ref{ubparams} we plot the fraction of blue/red and high/low S\'{e}rsic index as a function of redshift. In the right hand panel we split the sample into blue/red and high/low asymmetry. The errors are from Monte Carlo analysis as described in Section \ref{sec:sfr}

From Figure \ref{ubfrac} and \ref{ubparams} we find definite links between colour and visual morphology and colour and structure. We find that the reddest galaxies are generally spheroidal, with high S\'{e}rsic indices and low asymmetries. The bluest galaxies are generally peculiar/disturbed or forming disks and have low S\'{e}rsic indices. Also the most asymmetrical galaxies also trend to be blue (we note that the total fraction of high asymmetry galaxies is low compared with our visual peculiar fraction but this is to be expected, as the galaxies with the highest asymmetries are the galaxies at the most disturbed phase of a merger). This coupled with Figure \ref{sersicsfr} suggests there is a correlation between colour and star formation and whether or not a galaxy has some signiture of formation across the redshift range $z=1-3$.

These results agree with several findings in the literature. For example, studies such as \citet{Wein11},  \citet{Yuma11} and \citet{Bell12}, show that S\'{e}rsic index and the star-formation rate correlate strongly. In this work we extend this link to include colour. Furthermore, \citet{Wang12} show that how concentrated a galaxy is (as measured by Gini/M20) correlates well with star formation. Both \citet{Bell12} and \citet{Wang12} suggest passiveness is linked with bulge formation, which is in good agreement with the links found in this work. However, links between colour and structure are complicated by the presence of objects such as quenched or dusty disk galaxies which appear red in $U-B$ but will have disky or peculiar visual morphologies and extended S\'{e}rsic indices. For example, both \citet{Bell12} and \citet{Bruc12} note that there are quiescent galaxies which have a prominent disk component.

Overall we find several correlations between various structural parameters and morphology. However, there are often some mismatches between visual morphology, CAS morphology and S\'{e}rsic index. This is most likely due to these measures of morphology tracing different aspects of galaxy structure. S\'{e}rsic index traces how extended a galaxy is, but it tells us little about how disturbed a galaxy is. Visual morphology is sensitive to high surface brightness features and disturbances, and so is very good at distinguishing between galaxies which are smooth and galaxies which show signs of activity, such as mergers or star formation (e.g. spiral arms in the local Universe). CAS parameters can trace both of these, however, we cannot apply selections which are calibrated to local galaxies as the high redshift Universe is too structurally distinct. All of these parameters are important as they trace the star formation history of galaxies but we need to understand how to use them to best describe the high diversity of morphology at high redshift. We suggest that to fully explain visual morphology, at the redshifts discussed, a much more sophisticated visual classification scheme combined with structural parameters is needed to adequately classify galaxies at this epoch of galaxy formation.

\section{Summary}
We visually classify and study the star forming, colour and structural properties of a sample of 1188 galaxies with $M_{*}\ge10^{10}M_{\odot}$ and $z=1-3$. We calculate the fraction of galaxies of a given morphological type as a function of redshift and stellar mass. We also examine how our visual classifications compare to S\'{e}rsic index, $U-B$ colour and star-formation rate and we conclude the following:
\begin{itemize}

\item We find that the Universe at $z>2$ is dominated by peculiar galaxies, although there is still a substantial spheroid population which suggests the formation mechanisms of the Hubble sequence are already present at this epoch. We find that the Universe is not dominated by the types of morphologies we see in the local Universe until $z_{trans}=1.86\pm0.62$.

\item We investigate the influence that misclassification, due to image problems, may have on our results through simulations. We find that using our method we have a tendency to overestimate the spheroid fraction, and underestimate the peculiar fraction. However,the corrected fractions which we calculate in this work are generally within the errors of the uncorrected fractions, and the correction does not change the overall results found.

\item We examine the morphological fractions split into different stellar mass bins and find there is a dependence on the emergence of the Hubble sequence with stellar mass whereby the morphologies of the most massive galaxies have become settled earlier  in the life of the Universe than the less massive galaxies.

\item We find a negligible morphologically selected visual disk fraction at $z>2$. This is at odds with results from previous studies which have found that low S\'{e}rsic indices galaxies dominate the high redshift Universe. We suggest that this is a consequence of two effects. Firstly, disk galaxies at high redshift are in formation and/or being disturbed by other processes such as mergers. This leads to these galaxies being visually classified as peculiar galaxies. This results in a low fraction of Hubble type morphologically selected disks. Secondly, peculiar galaxies are often extended and have low S\'ersic indices. They therefore contribute to the high fraction of galaxies with low S\'{e}rsic indices, and hence low S\'ersic index is not an indication of a disk galaxy at high redshift.

\item We look at the evolution of the number density of each galaxy type and find that the number density of the peculiar galaxies remains fairly constant across the redshift range $z=1-3$, but the ellipticals and disks increase with time at a roughly constant rate from $z=1-3$.

\end{itemize}
We also find links between visual morphology, rest frame colour and star formation rate. There is a correlation between how extended an object is, how disturbed it is and the colour and star formation rate of the galaxy, such that peculiar and disturbed disk type galaxies are bluer and have higher specific star formation. To understand these connections further at high redshift, we need to investigate a larger sample which contains more galaxies at various phases of formation.

Another major issue discussed within this paper is that we investigate galaxy morphology using a simple classification system based on those developed for the $z=0$ Universe. However, as this paper shows, this morphological classification system breaks down at high redshifts. This is not, as previously assumed, because galaxies are all peculiar at $z > 1$, as they are not, but because many systems look smooth and compact, like ellipticals, but have high star formation rates and blue colours. The great diversity of these galaxies makes classifying them by morphology or other properties a non-trivial task. Although there are some galaxies present which resemble the Hubble type galaxies we see in the local Universe, the presence of disturbed forming disks, dusty galaxies, and compact red and blue objects causes discord in defining morphology or in general galaxy classification. When higher resolution imaging is available a classification scheme may naturally present itself. Until then perhaps a combination of structure and physical properties is needed to classifying forming galaxies at $z > 1$.  In 
the future the James Webb Space Telescope (JWST) and Euclid will provide a further advance for this, although ultimately a high resolution telescope in the NIR which exceeds that provided by WFC3 is needed. Furthermore, the understanding of galaxy structure can be advanced by exploring the kinematics of galaxies at high redshift. Future Integral Field Units (IFUs), such as the K-band Multi-Object Spectrograph (KMOS) on the Very Large Telescope (VLT), will allow for observations of multiple objects and we can obtain kinematic information for large numbers of galaxies.

\section*{Acknowledgments}
We would like to thank the CANDELS team for their support and work on the survey and this paper. We would like to thank Casey Papovich, Ray Lucas and Yu Lu for their helpful discussion. We would also like to acknowledge funding from the STFC and the Leverhulme Trust.

\bibliographystyle{mnras}
\bibliography{refs}
\label{lastpage}
\end{document}